\documentclass[pra,showpacs,twocolumn]{revtex4}

\usepackage{graphicx}
\usepackage{amsmath}
\usepackage{subfigure}
\usepackage{epsfig}

\begin{document}

\title{\bf Coherent control of ac Stark allowed transition in $\Lambda$ system}

\author{Gennady A. Koganov}
\email{quant@bgu.ac.il}
\affiliation{Physics Department, Ben Gurion University of the Negev\\
P.O.Box 563, Beer Sheva 84105, Israel}
\author{Reuben Shuker}
\email{shuker@bgu.ac.il}
\affiliation{Physics Department, Ben Gurion University of the Negev\\
P.O.Box 563, Beer Sheva 84105, Israel}

\begin{abstract}
We show that quantum-interference-related phenomena, such as
electromagnetically induced transparency, gain without inversion and
enhanced refractive index may occur on electric-dipole
forbidden transitions. Gain/dispersion characteristics of
such transitions strongly depend upon
the relative phase between the driving and probe fields. Unlike allowed transitions, gain/absorption behavior of forbidden transitions exhibit antisymmetric feature on the Rabi sidebands. Absorption/gain spectra possess extremely narrow sub-natural resonances.
\end{abstract}


\maketitle

Quantum-interference-related phenomena, such as electromagnetically
induced transparency (EIT), lasing without inversion (LWI) etc., are
based on the interference between two independent quantum channels
\cite{LWI,EIT-review,Agarwal,LWI-in-Na,LWI-review,Quantum-control,Arimondo,Velichan,Maser,Rb-maser}.
Traditional treatment of such phenomena typically involves a three
level scheme and two coherent fields, a strong driving field and a
weak probe one, applied to two dipole-allowed transitions, followed
by measurement of the absorption and the dispersion of the probe
transition.

In this Letter we show that dipole-forbidden transitions can also exhibit amplification, enhanced dispersion and other coherent phenomena. The electric field component of the driving laser field breaks the space spherical symmetry and renders the parity not well defined, as in \textit{dc} Strark effect. In other words, the presence of a strong driving field exerts \textit{ac} Strark effect and thus breaks the spherical symmetry of the system and creates an infinite ladder of dressed states \cite{CohenTanudji,Doron2001}. An interesting signature of the forbidden transitions is the antisymmetric character of the gain and dispersion on the Rabi sidebands. We found that gain and dispersion properties of such transitions are phase sensitive as they strongly depend upon the relative phase between the driving and probe fields. Our results open a perspective for new type of phase sensitive spectroscopy in a wide spectral range.

\begin{figure}[h]
\centering
\includegraphics[scale=0.5]{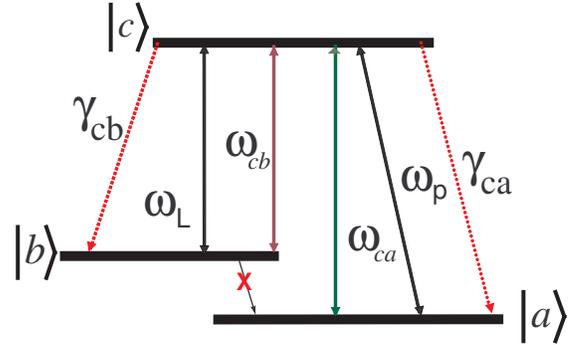}
\caption{Schematic $\Lambda$ three level
configuration. Transition
$\left|a\right\rangle\rightarrow\left|b\right\rangle$ is initially dipole-forbidden (marked with red cross).} \label{Fig1}
\end{figure}

We consider a tree level scheme in $\Lambda$-configuration shown in Fig. \ref{Fig1}.  A strong driving field with Rabi frequency
$\Omega_{L}exp(i\varphi_{L})$ and a weak probe field with Rabi frequency
$\Omega_{P}exp(i\varphi_{P})$ are applied to atomic
transitions $\left|b\right\rangle\rightarrow\left|c\right\rangle$
and $\left|a\right\rangle\rightarrow\left|c\right\rangle$, respectively.
Note that the phases $\varphi_{L}$ and $\varphi_{P}$ are introduced
explicitly in the driving and the probe fields, respectively. 

Due to selection rules the transition $\left|a\right\rangle\rightarrow\left|b\right\rangle$ is dipole-forbidden by parity which is well defined in the absence of the laser fields. However, the parity becomes ill defined due to the presence of the ac electric fields of the two impinging lasers. These fields break the space symmetry and mix states of different parity. Hence the originally forbidden transition becomes undefined. In this sense it becomes \textit{dynamically} allowed. We call such dynamically allowed transitions ac-Stark allowed (ACSA) transitions. In other words the interaction with two coherent fields, at least one of which is strong, creates an infinite ladder of
dressed states \cite{CohenTanudji,Doron2001}, so that transitions
between the dressed states are not necessarily constrained by the
selection rules for a free atomic system. In a sense this is a
different variant than the usual EIT or LWI schemes as here the
transition is forbidden to begin with and there is no issue of
population inversion. The transition is dynamically allowed solely due to the
quantum interference.
\bigskip
The system Hamiltonian in the interaction picture and the rotating wave approximation is given by

\begin{figure}[h]
\begin{center}
\includegraphics[scale=0.6]{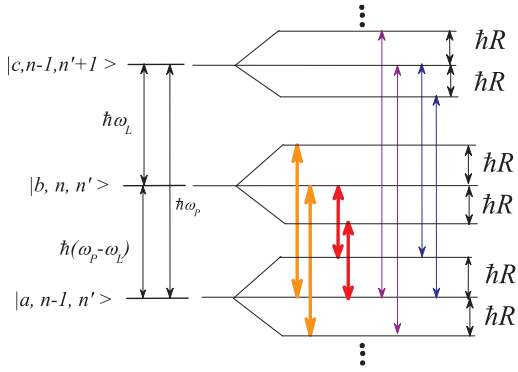}
\end{center}
\caption{Dressed states picture at bare two-photon resonance
$\Delta_{P}-\Delta_{L}=0$. On the left: manifold of bare states labeled
by the atomic level index \textit{a}, \textit{b}, or \textit{c},
the driving field photon number \textit{n}, and the probe field
photon number \textit{n'}.  On the right: dressed states of coupled
atom+field system. Strong phase sensitive gain
without inversion is obtained on the ACSA transition
$\left|a\right\rangle\rightarrow\left|b\right\rangle$ at
frequencies $\omega_{P}-\omega_{L}+R$, left couple of thick arrows (orange on line) and $\omega_{P}-\omega_{L}-R$, right couple of thick arrows (red on line), see also Fig. \ref{r01=Pi}. Two couples of thin arrows(blue and violet on line), show the probe transition frequencies $\omega_{P}-R$ and
$\omega_{P}+R$, where gain is also possible.
$R=\sqrt{\Omega_{P}^{2}+\Omega_{L}^{2}}$ is the generalized Rabi
frequency.} \label{DressedStates}
\end{figure}

\begin{equation}\label{Hamiltonian}
H=\left(
\begin{array}{lll}
 0 & \Omega_{L}e^{i\varphi_{L}} & \Omega_{P}e^{i\varphi_{P}} \\
\Omega_{L}^{*}e^{-i\varphi_{L}} & \Delta_{L} & 0 \\
\Omega_{P}^{*}e^{-i\varphi_{P}} & 0 & \Delta_{P}
\end{array}
\right)
\end{equation}

\noindent where $\Omega_{P}$ and $\Omega_{L}$ are Rabi frequencies of the probe and the driving fields, respectively, and $\Delta_{P}$ and $\Delta_{L}$ are corresponding detunings.  Abbreviated manifold of dressed states created by the strong driving field $\Omega_{L}$ is
shown in Fig. \ref{DressedStates}. As will be seen in the following,
maximal gain is achieved when the probe and the drive lasers are in
''dressed'' two-photon resonance with transitions between the
dressed states marked with thick arrows (red and orange on line), i.e. at
$\Delta_{P}-\Delta_{L}=\pm R$, where $R=\sqrt{\Omega_{P}^{2}+\Omega_{L}^{2}}$. The corresponding semiclassical dressed states in the case of resonant drive field are given by

\begin{equation}
\left|\pm\right\rangle=\frac{1}{\sqrt{2}}(\frac{\Omega_{P}}{\sqrt{\Omega_{P}^{2}+\Omega_{L}^{2}}}\left|a\right\rangle +\frac{e^{i\Delta\varphi}\Omega_{L}}{\sqrt{\Omega_{P}^{2}+\Omega_{L}^{2}}}\left|b\right\rangle \pm e^{-i\varphi_{P}}\left|c\right\rangle)
\end{equation}
 
\noindent These dressed states are superposition of states of different parity and hence are not constrained by the selection rules for atomic states $\left|a\right\rangle, \left|b\right\rangle$, and $\left|c\right\rangle$.

We have solved analytically master equation
for the atomic density matrix $\rho$ with the Hamiltonian (\ref{Hamiltonian}) in steady state and calculated the coherences $\rho_{ac}$ and $\rho_{ab}$ on the probe and the ACSA transitions,
whose imaginary and real parts are related to absorption/gain and
dispersion, respectively. The main features can be qualitatively
understood from the approximate expressions for the coherences
$\rho_{ac}$ and $\rho_{ab}$ on probe and ACSA transitions at
small probe field $\Omega_{P}\ll\Omega_{L}$, although all graphical
results presented in the following have been obtained from the
exact analytic solution of the steady state master equation. 
Approximate formulas for the coherences are given by

\begin{equation}
\rho_{ac}=-\frac{2i(\Delta_{L}-\Delta_{P})\Omega_{P}e^{i\varphi_{P}}}{(\gamma_{ca}+\gamma_{cb})(\Delta_{L}-\Delta_{P})+2i(\Delta_{P}(\Delta_{L}-\Delta_{P})+\Omega_{L}^{2})} \label{r02}
\end{equation}
\begin{equation}
\rho_{ab}=\frac{2i\Omega_{P}\Omega_{L}e^{i(\varphi_{P}-\varphi_{L})}}{(\gamma_{ca}+\gamma_{cb})(\Delta_{L}-\Delta_{P})-2i(\Delta_{P}(\Delta_{L}-\Delta_{P})+\Omega_{L}^{2})}\label{r01}
\end{equation}

\noindent Here $\gamma_{ca}$ and $\gamma_{cb}$ are spontaneous decay
rates from the upper state $\left|c\right\rangle$ to states
$\left|a\right\rangle$ and $\left|b\right\rangle$, respectively,
$\Delta_{P}=\omega_{ac}-\omega_{P}$ and
$\Delta_{L}=\omega_{bc}-\omega_{L}$ are one-photon detunings between
the laser frequencies $\omega_{P}$ and $\omega_{L}$ and the
corresponding atomic frequencies $\omega_{ac}$ and $\omega_{bc}$.
The two-photon detuning
$\Delta_{P}-\Delta_{L}=\omega_{ab}-(\omega_{P}-\omega_{L})$ contains
the frequency $\omega_{ab}$ of originally dipole-forbidden atomic
transition $\left|a\right\rangle\rightarrow\left|b\right\rangle$
indicating possible oscillations at that frequency. It is important to note that
due to the presence of the exponent factor
$exp[i(\varphi_{P}-\varphi_{L})]$ in the numerator of Eq.
(\ref{r01}), the coherence $\rho_{ab}$ is \textit{phase sensitive}. Varying the relative phase
$\Delta\varphi=\varphi_{P}-\varphi_{L}$ between the probe and
 the drive fields
allows for an interchange between absorptive and dispersive line shapes of
$\left|a\right\rangle\rightarrow\left|b\right\rangle$ transition, defined by imaginary and real parts of the coherence $\rho_{ab}$,
respectively. It is instructive to consider two particular cases
of "bare" and "dressed" two-photon resonances, when
$\Delta_{P}=\Delta_{L}$ and $\Delta_{L}-\Delta_{P}=\pm
\Omega_{L}$,. At $\Delta_{P}=\Delta_{L}$ ("bare
two-photon resonance") the coherences become $\rho_{ac}=0$ and $\rho_{ab}=-(\Omega_{P}/\Omega_{L})exp(i\Delta\varphi)$, so that for transition $\left|a\right\rangle\rightarrow\left|b\right\rangle$ both absorption and dispersion can take
small values between $-\Omega_{P}/\Omega_{L}$ and
$\Omega_{P}/\Omega_{L}$, depending on the relative phase $\Delta\varphi$. This accentuates the importance of the phase shift between the two laser fields and the phase sensitivity of the ACSA transition. The absorption $\rho_{ac}$ on the probe transition $\left|a\right\rangle\rightarrow\left|c\right\rangle$ equals
to zero, the signature of EIT and dark state\cite{LWI,EIT-review}.

A different picture is obtained when the probe field is tuned in
resonance with transitions between dressed states, i.e. when the two
photon detuning equals the Rabi frequency of the driving field ("dressed two-photon
resonance"). The dressed resonance condition and the coherences are given by

\begin{align}
\Delta_{L}-\Delta_{P}=\pm \Omega_{L} \label{dressed-res}\\
\rho_{ac}=-\frac{2i\Omega_{P}}{{\gamma_{cb}+\gamma_{ca}+2i\Delta_{L}}}e^{i\varphi_{P}} \label{r02-res-dress}\\
\rho_{ab}=\pm\frac{2 i \Omega_{P}
e^{i\Delta\varphi}}{\gamma_{cb}+\gamma_{ca}-2i\Delta_{L}} \label{r01-res-dress}
\end{align}

\noindent In this case both $\rho_{ac}$ and $\rho_{ab}$ are not small since they
are proportional to the probe Rabi frequency $\Omega_{P}$ rather
than to the ratio $\Omega_{P}/\Omega_{L}$ between the probe and the
driving field Rabi frequencies, as in the above mentioned bare resonance case. Also gain and dispersion on the ACSA
transition are not symmetric with respect to the bare two-photon
resonance $\Delta_{P}=\Delta_{L}$, as evident from Eq. (\ref{r01-res-dress}). 
In particular, at zero phase
shift $\Delta\varphi=0$ and zero drive detuning $\Delta_{L}=0$ there is strong gain with
anomalous dispersion at $\Delta_{P}=-\Omega_{L}$ and strong
absorption with normal dispersion at $\Delta_{P}=\Omega_{L}$.
Changing the phase shift to $\Delta\varphi=\pi$ results in an
opposite sign of $\rho_{ab}$, so that there is absorption with normal dispersion at
$\Delta_{P}=-\Omega_{L}$ and gain with
anomalous dispersion at $\Delta_{P}=\Omega_{L}$.

\begin{figure}[h]
\centering
\subfigure[]{\label{r01=0}\includegraphics[scale=0.35]{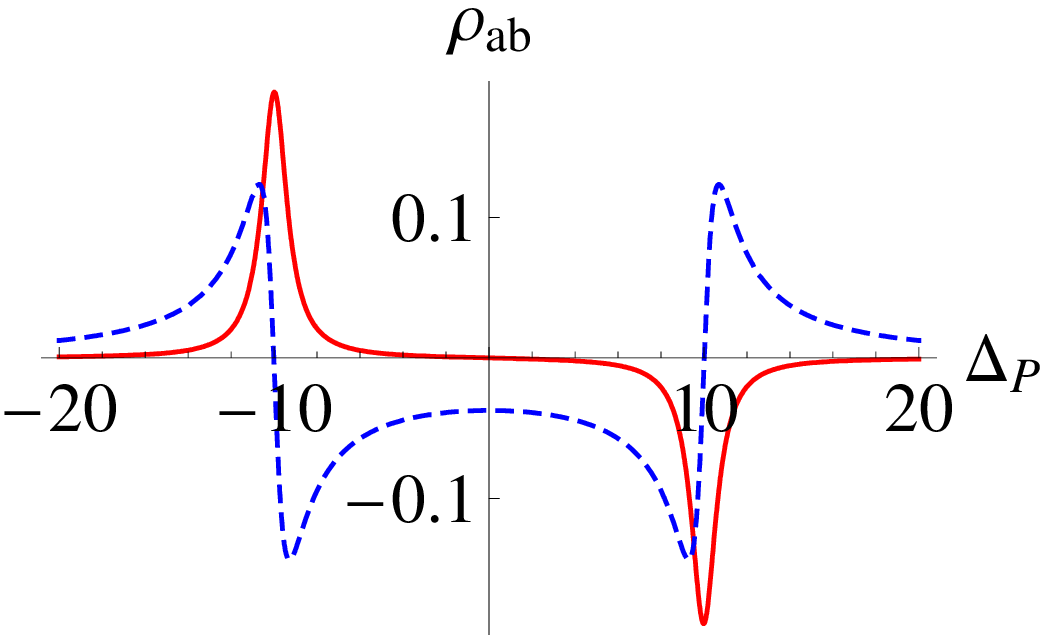}}
\subfigure[]{\label{r01=0.5Pi}\includegraphics[scale=0.35]{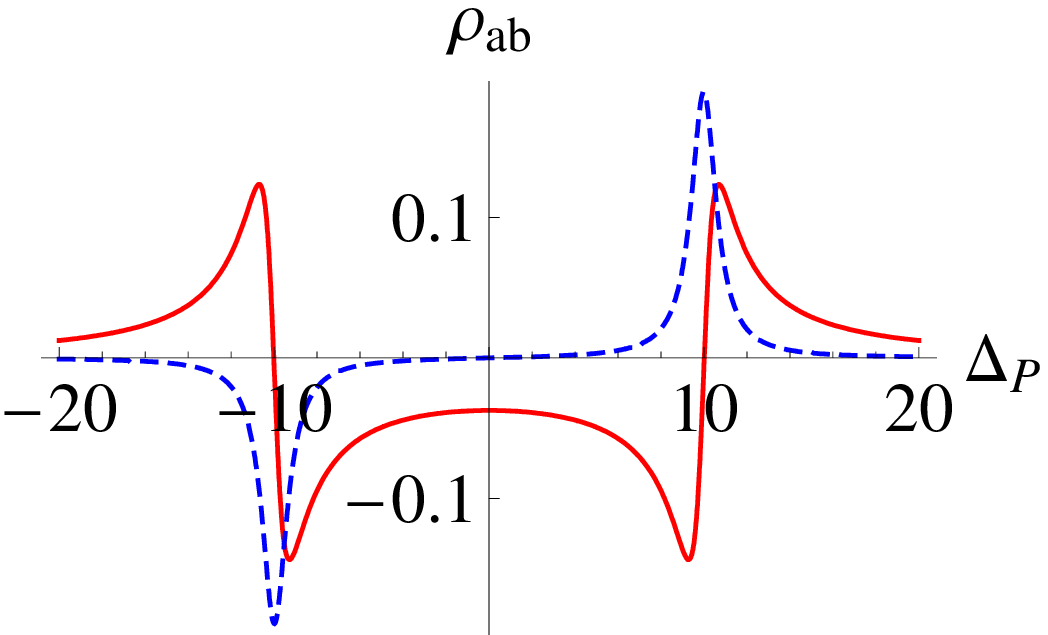}}\\
\subfigure[]{\label{r01=Pi}\includegraphics[scale=0.35]{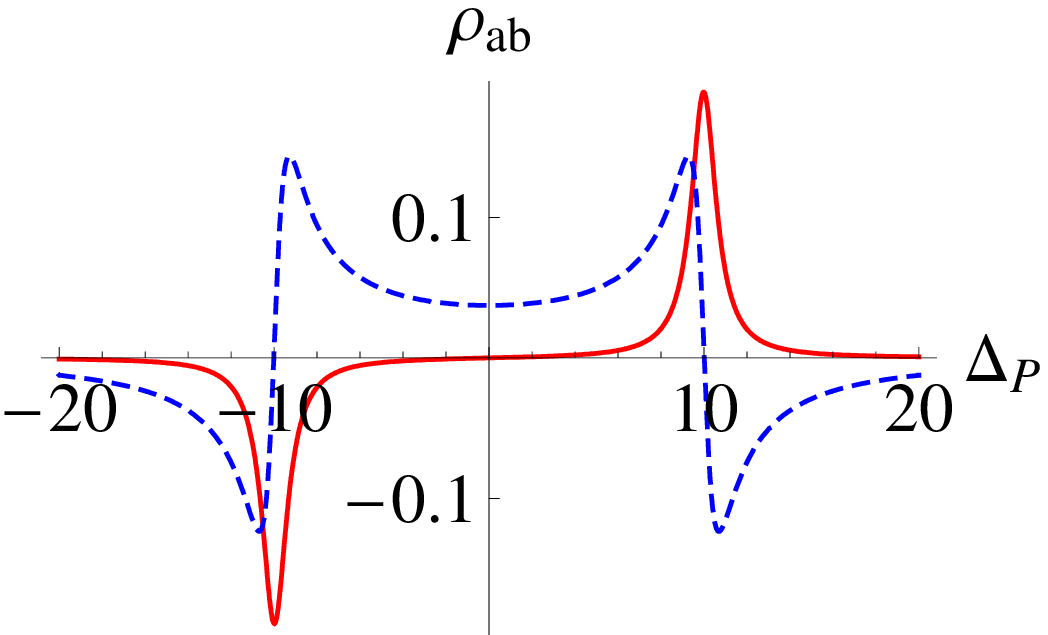}}
\subfigure[]{\label{r01=1.5Pi}\includegraphics[scale=0.35]{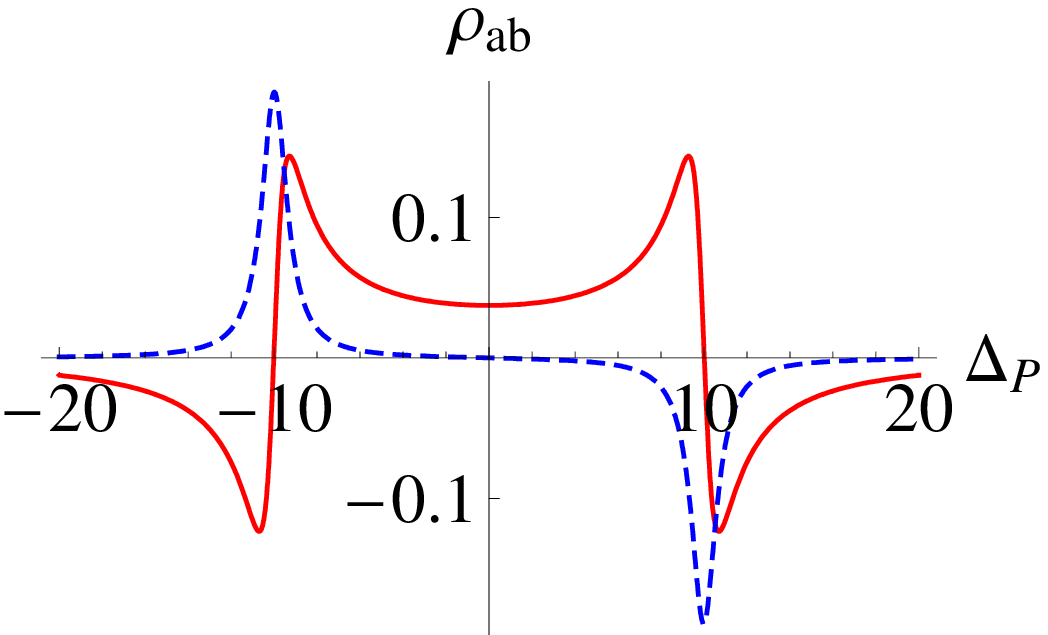}}\\
\subfigure[]{\label{Forbid-coherence-map}\includegraphics[scale=0.35]{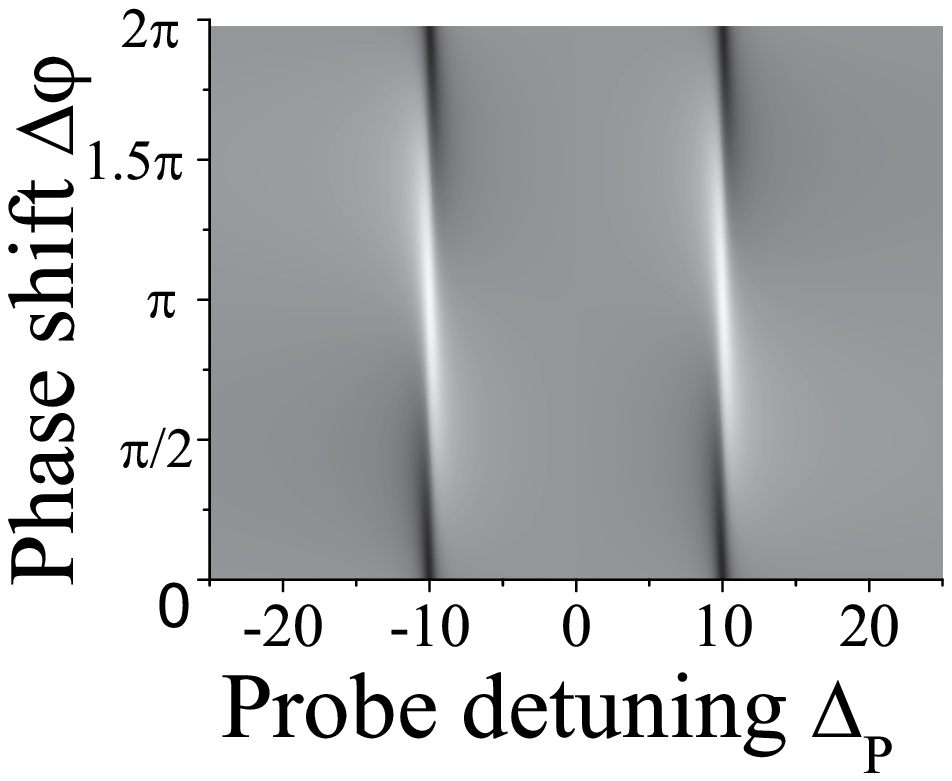}}
\subfigure[]{\label{Probe-coherence-map}\includegraphics[scale=0.35]{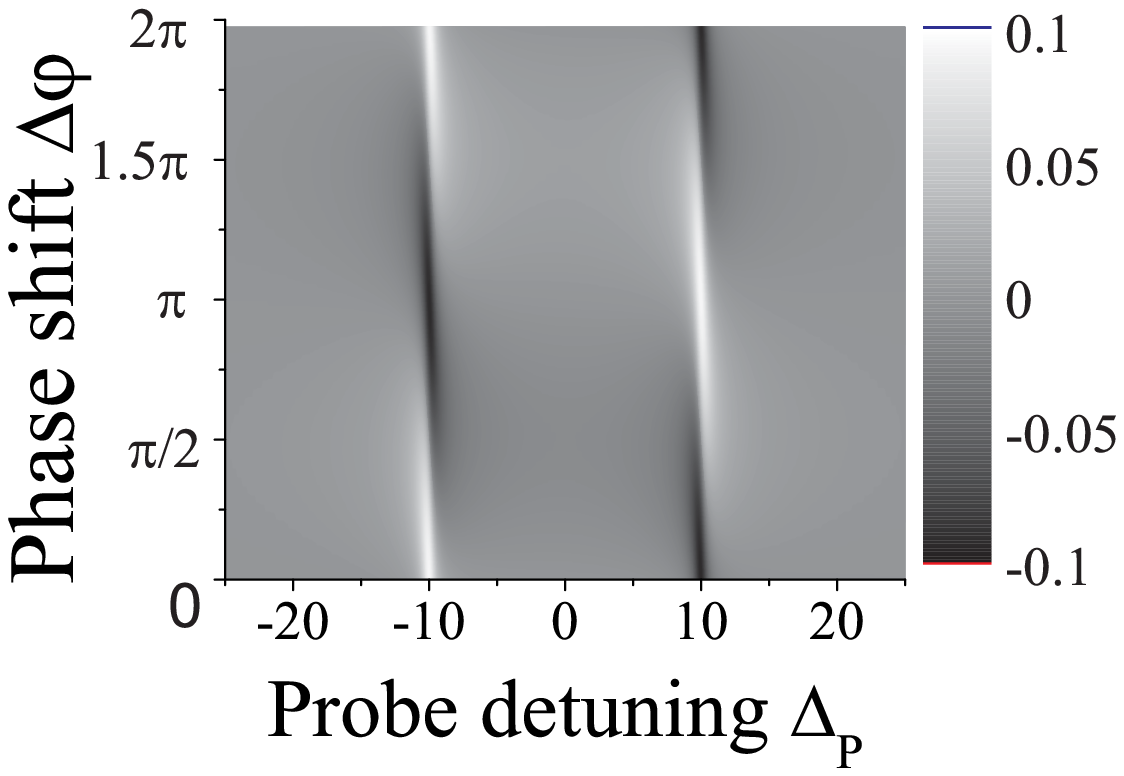}}
\caption{$Im[\rho_{ab}]$ (solid curve, red on-line) and $Re[\rho_{ab}]$ (dashed
curve, blue on-line) as functions of probe detuning at (a) $\Delta\varphi=0$, (b)
$\Delta\varphi=\pi/2$, (c) $\Delta\varphi=\pi$, and (d)
$\Delta\varphi=3\pi/2$. The antisymmetric behavior of the gain is accompanied by symmetric behavior of the dispersion at specific phase shifts, and vise versa. Parts (e) and (f) show contour plots of
steady-state coherences $\rho_{ac}$ and $\rho_{ab}$ at probe and
ACSA transitions, respectively. White/black color corresponds
to maximal gain/absorption. Parameters: $\gamma_{ca}=\gamma_{cb}$,
$\Omega_{L}=10\gamma_{cb}$, $\Omega_{P}=0.37\gamma_{cb}$.}
\label{Lambda-Forbid-Phase}
\end{figure}

To get a quantitative picture, exact analytical solution for the steady state
coherence $\rho_{ab}$ is drawn in Fig. \ref{Lambda-Forbid-Phase} as
a function of probe detuning $\Delta_{P}$ for
various values of $\Delta\varphi$.
When the probe and the driving fields are \textit{in-phase} ($\Delta\varphi=0$,
see Fig. \ref{Lambda-Forbid-Phase}a) there is a strong gain with
negative dispersion slope at $\Delta_{P}=-\Omega_{L}$, and
absorption with normal dispersion at $\Delta_{P}=\Omega_{L}$.
Inverse picture is obtained when the probe and driving fields are
\textit{out of phase}, i.e. $\Delta\varphi=\pi$ (see Fig.
\ref{Lambda-Forbid-Phase}c). If the field phases are
$\pm\pi/2$-shifted (see Figs. \ref{Lambda-Forbid-Phase}b and
\ref{Lambda-Forbid-Phase}d), the absorption profile takes a
dispersive shape while the dispersion behaves in absorptive-like
manner - an indication of the quantum interference. To compare
absorption/gain properties of the probe and ACSA transitions
contour plots of imaginary parts of $\rho_{ac}$ and $\rho_{ab}$
are shown in Figs. \ref{Lambda-Forbid-Phase}e and
\ref{Lambda-Forbid-Phase}f, respectively. One can clearly see the
difference: the probe coherence $\rho_{ac}$ is symmetric with
respect to bare two-photon resonance $\Delta_{P}=\Delta_{L}$ so that
there is either absorption or gain simultaneously on both side-bands
$\Delta_{P}-\Delta_{L}=\pm\Omega_{L}$ (two white traces at
$\Delta\varphi=\pi$ and two black traces at
$\Delta\varphi=0(2\pi)$), while the coherence $\rho_{ab}$ on the ACSA transition is
antisymmetric with respect to the origin, so that there is always gain on one side-band and
absorption on the other (white trace on one side-band is always
accompanied by the black trace on the opposite side-band).
The frequency at which amplification takes place can be
tuned by varying the driving field intensity because maximal gain
is obtained at $\Delta_{P}-\Delta_{L}=\pm\Omega_{L}$.

\begin{figure}[h]
\begin{center}
\subfigure[]{\label{RRI}\includegraphics[scale=0.35]{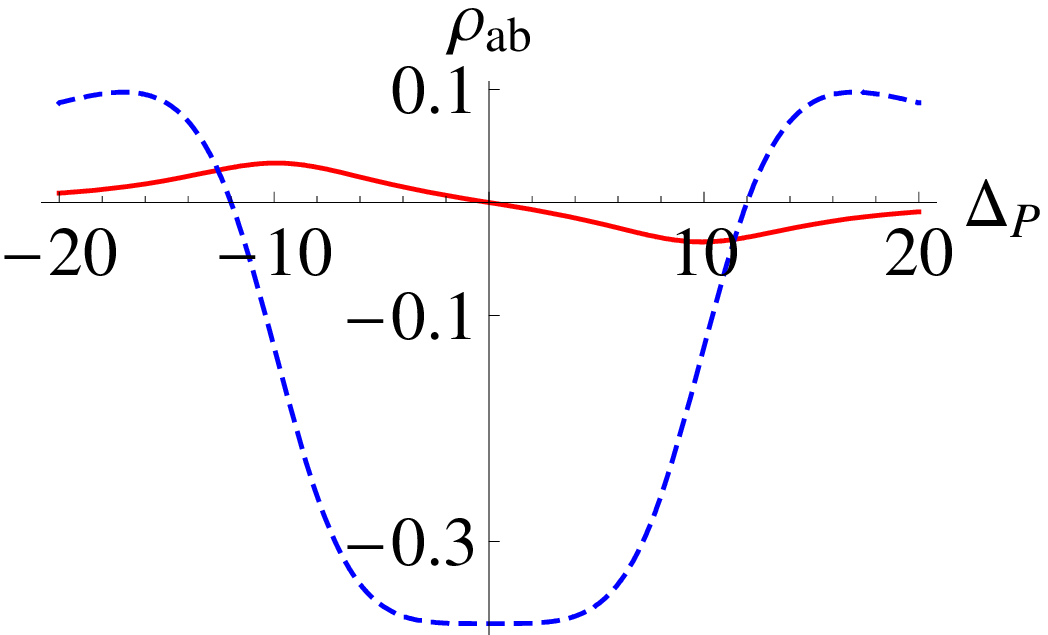}}
\subfigure[]{\label{AWI}\includegraphics[scale=0.35]{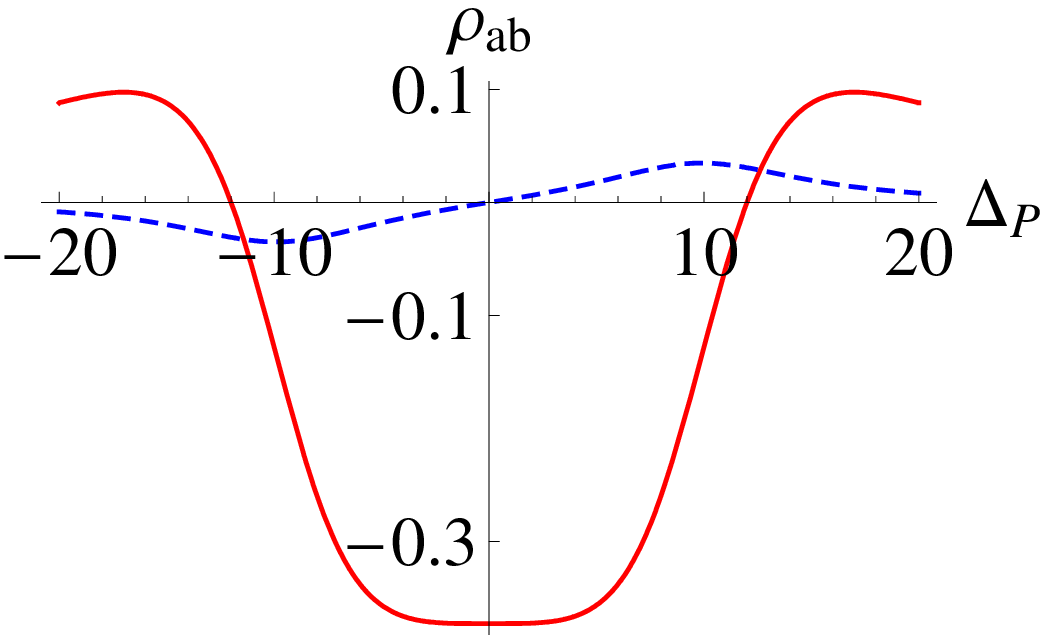}}\\
\subfigure[]{\label{ERI}\includegraphics[scale=0.35]{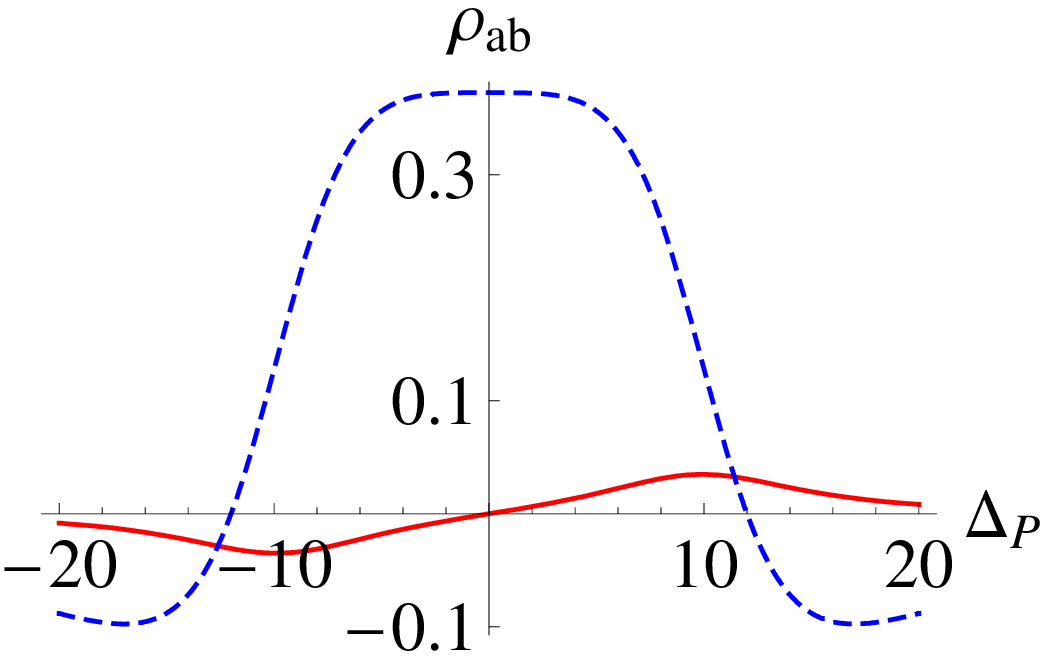}}
\subfigure[]{\label{GWI}\includegraphics[scale=0.35]{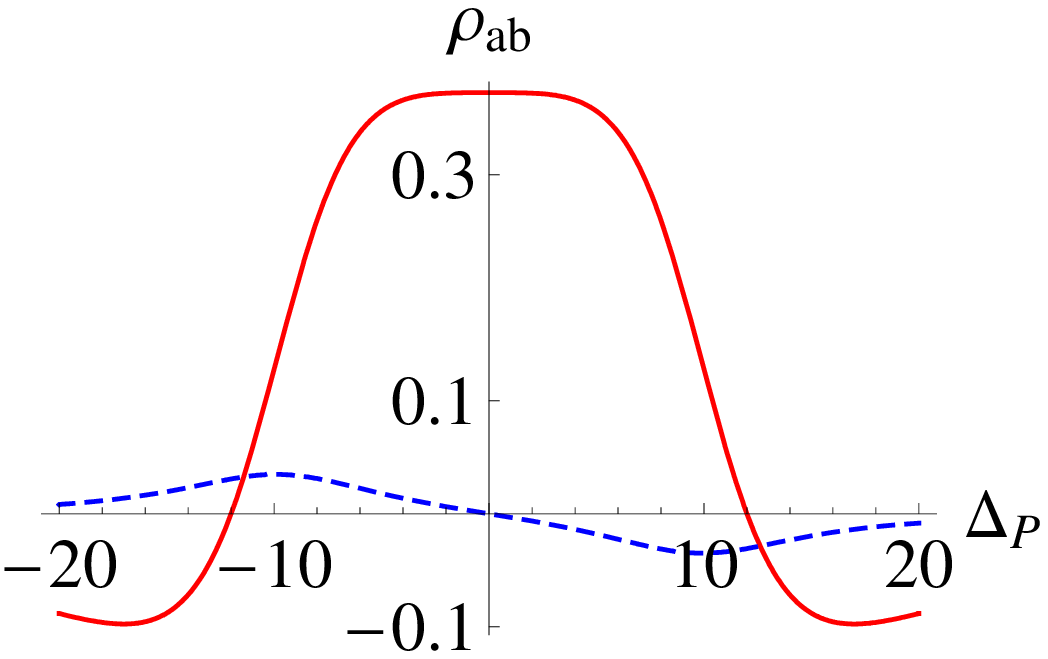}}\\
\subfigure[]{\label{Absorp-plato}\includegraphics[scale=0.8]{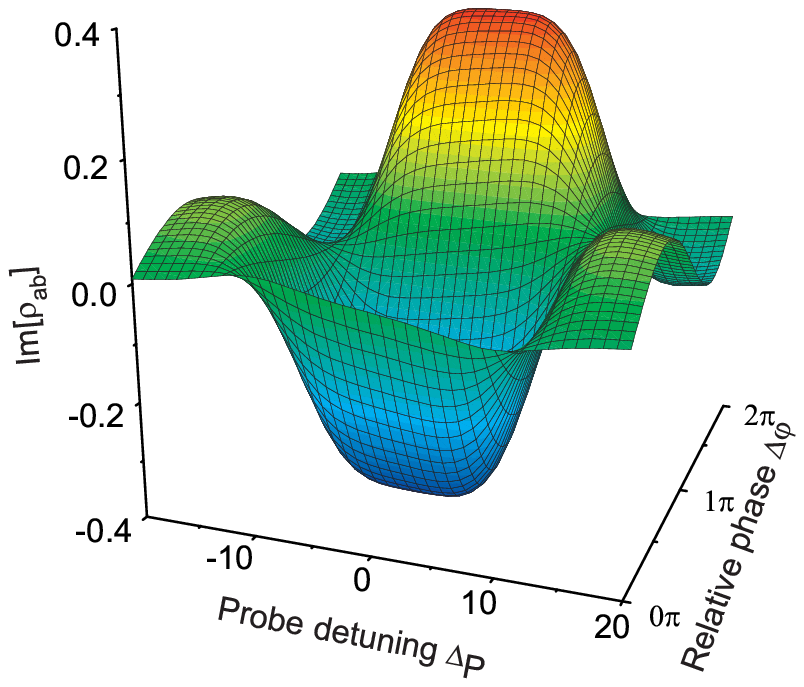}}
\end{center}
\caption{Imaginary (solid curve, red on-line) and real (dashed curve, blue on-line) parts of the
coherence $\rho_{ab}$ on the ACSA transition at strong pump
field as functions of probe detuning. Gain (absorption) is obtained at positive (negative) values. (a) Suppressed refraction index
at $\Delta\varphi=0$, (b) Strong absorption with normal dispersion
at $\Delta\varphi=\pi/2$, (c) Enhanced refraction index at
$\Delta\varphi=\pi$,  (d) Gain without inversion with negative
dispersion slope at $\Delta\varphi=-\pi/2$. Figure (e) shows
3D plot of imaginary part of $\rho_{ab}$ as a function of the probe detuning $\Delta_{P}$ and of the phase shift $\Delta\varphi$. The dispersion behavior is similar and can be obtained by shifting the gain plot by $\pi/2$. Note the unique feature of flat-top gain and dispersion (not shown here) at particular values of the phase shift $\Delta\varphi$. Parameters:
$\gamma_{ca}=\gamma_{cb}$, $\Omega_{L}=10\gamma_{cb}$,
$\Omega_{P}=4.5\gamma_{cb}$.} \label{EnhancedRefrac}
\end{figure}

Another intriguing manifestation of quantum interference is obtained
when the probe field is not weak with respect to the driving one.
Figure \ref{EnhancedRefrac} demonstrates four essentially different
cases controlled by the \textit{phase shift} between the probe and
driving fields: (i) reduced refraction index at $\Delta\varphi=0$
(fields in-phase, Fig.\ref{RRI}), (ii) strong absorption with normal dispersion at
$\Delta\varphi=\pi/2$ (Fig.\ref{AWI}), (iii) strongly enhanced refraction index at
$\Delta\varphi=\pi$ (out-of-phase fields, (Fig.\ref{ERI})), and (iv) strong gain
without inversion and anomalous dispersion at
$\Delta\varphi=-\pi/2$ (Fig.\ref{GWI}). The results shown in Figs.
\ref{ERI} and \ref{GWI} are especially
important as they hint for two interesting potential applications:
laser or/and optical amplifier with a wide spectral range of
operation (see Fig. \ref{GWI}), and a controllable
atomic dispersion in wide spectral range (see Fig.
\ref{ERI}). Moreover, as one can observe form 3D
plot of the imaginary part of the coherence
$\rho_{ab}$ (see Fig. \ref{Absorp-plato}), the unique broad flat-top gain takes place in a wide range of phase shift
$\Delta\varphi$ as well.

\begin{figure}[htbp]
\centering
\subfigure[]{\label{r02=-20}\includegraphics[scale=0.35]{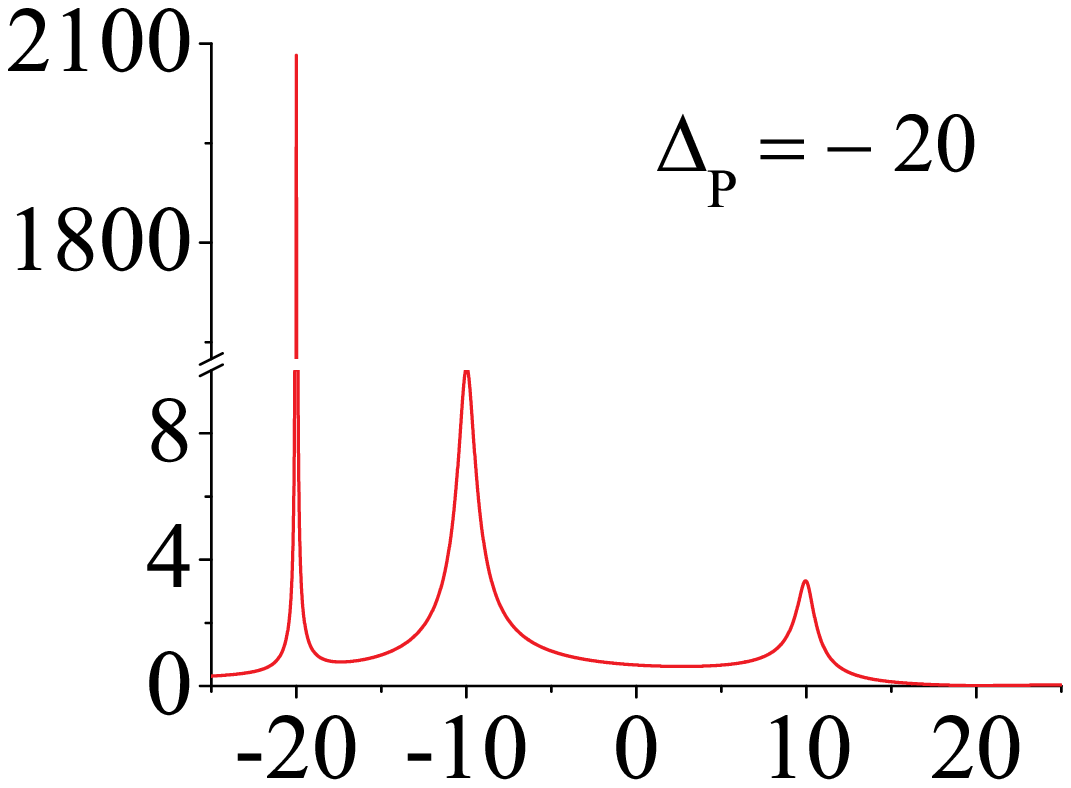}}
\subfigure[]{\label{r02=-15}\includegraphics[scale=0.35]{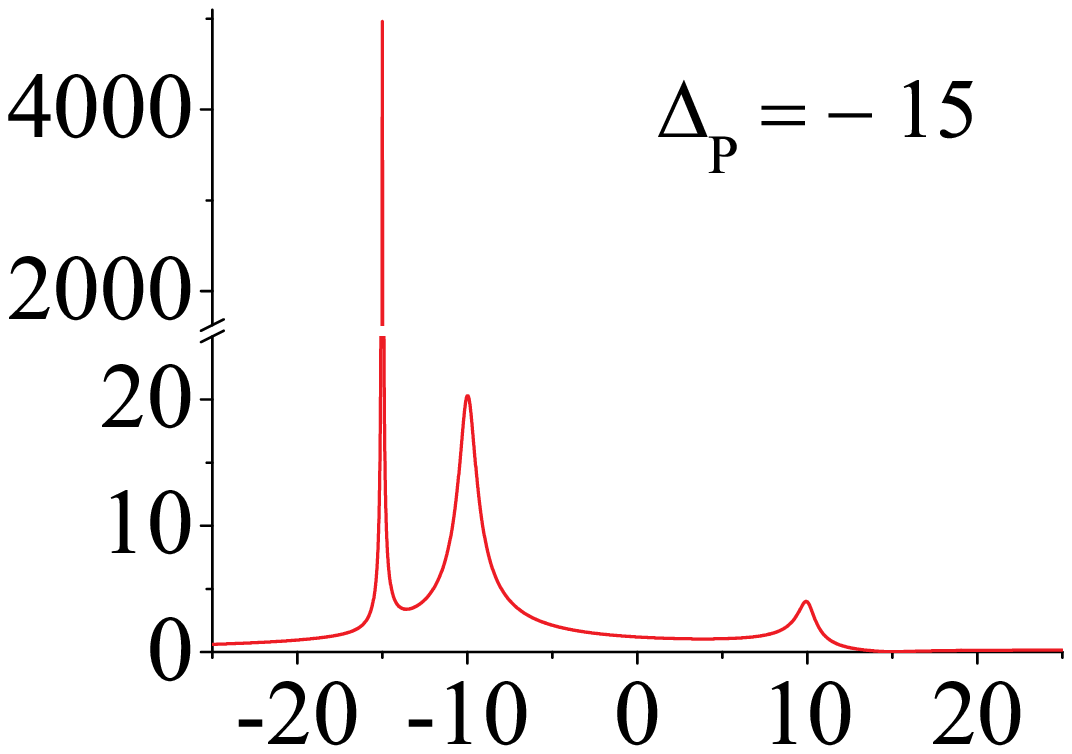}}\\
\subfigure[]{\label{r02=-10}\includegraphics[scale=0.35]{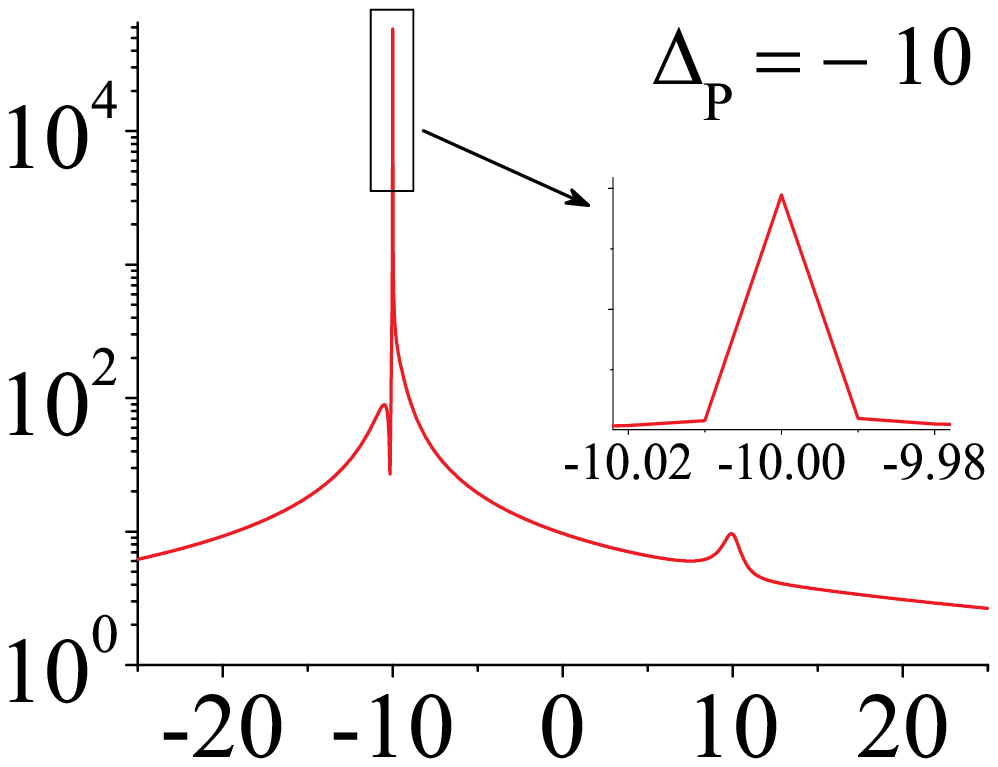}}
\subfigure[]{\label{r02=0}\includegraphics[scale=0.35]{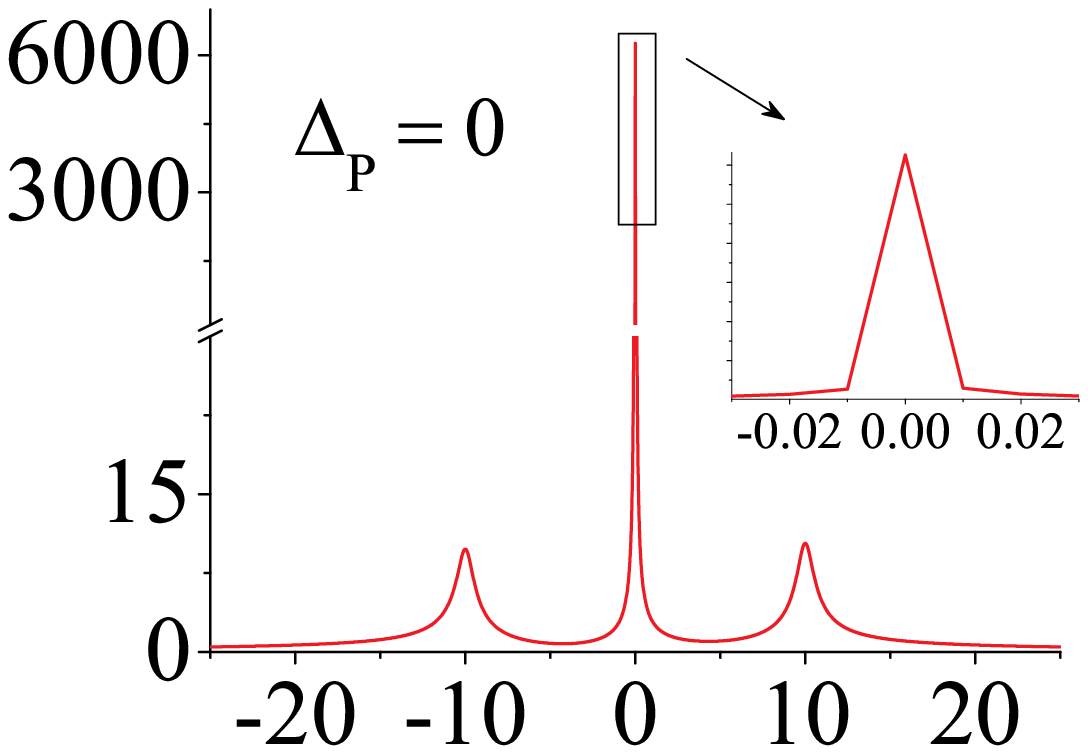}}
\caption{Absorption spectrum of
$\left|a\right\rangle\rightarrow\left|b\right\rangle$ transition
(initially dipole-forbidden). Note sharp strong peak at dressed
two-photon resonance $\Delta_{P}=\pm\Omega_{L}$ (see part c which is plotted in
logarithmic scale).
Parameters: $\Omega_{L}=10, \Omega_{P}=0.1,
\Delta\varphi=0$, $\Delta_{L}=0$, $\Delta_{P}=-20$ (a), $-15$ (b), $-10$ (c), $0$ (d).} \label{LamForbidSpetr(deltaP)}
\end{figure}

To calculate the absorption spectrum, the time dependent master
equation has been solved numerically and then Fourier transform has
been taken. Figure \ref{LamForbidSpetr(deltaP)} plots the spectrum
of the coherence $\rho_{ab}$ for various values of the probe
detuning $\Delta_{P}$. As one can see, in addition to two resonances
at Rabi-shifted transition frequencies $\omega_{ab}\pm\Omega_{L}$, a
very sharp peak appears at $\omega_{ab}+\Delta_{P}-\Delta_{L}$. This
peak becomes especially strong by a few orders of magnitude, at the
''dressed'' two-photon resonance, i.e. at
$\Delta_{P}-\Delta_{L}=\pm\Omega_{L}$ (see Fig.
\ref{LamForbidSpetr(deltaP)}c).
Interestingly, the peaks have extremely narrow line widths (see
insets in Figs. \ref{LamForbidSpetr(deltaP)}c and \ref{LamForbidSpetr(deltaP)}d). We have found numerically that
increasing accuracy of digital Fourier transform gives rise to
vanishing line width, i.e. ideally, if one could perform Fourier
transform with infinite accuracy, the peaks would look like
$\delta$-function. This can be explained by the absence of
mechanisms of line-broadening on the forbidden transition in this
model. Indeed, there are neither dissipative processes which could
contribute to line width, nor the field broadening as no field is
applied to this transition. These will, of course, give a finite
line width.

Sharp strong resonance in the spectrum of
$\left|a\right\rangle\rightarrow\left|b\right\rangle$ transition
means that an efficient lasing is possible on this transition with
extremely narrow spectral line. Moreover, such a laser would be
tunable since its frequency can be controlled by varying both the
frequency of either the driving or the probe laser, and the
intensity of the driving laser.

It should be noted that our results are applicable to the systems
where all three transitions are electric dipole allowed, such as
Ruby and other solid materials \cite{Ruby}, and semiconductor
quantum wells \cite{Frog-Nature}.

Last but not least, similar results have been obtained for the
ladder and the V schemes as well. In particular, strong gain without
inversion can be obtained on dipole forbidden transition
$\left|a\right\rangle\rightarrow\left|c\right\rangle$ of the ladder
scheme (see Fig. \ref{Fig1}c) which opens a new perspective for
creating ultra-short wavelength and X-ray lasers without inversion.
Details will be published elsewhere.

   In summary, gain/absorption and dispersion characteristics of dipole-forbidden transitions in $\Lambda$ scheme driven by two laser fields have been calculated. These are found to be strongly sensitive to the phase shift between the probe and the drive laser fields. It has been shown that $\left|a\right\rangle\rightarrow\left|b\right\rangle$ transition exhibits quantum-interference-related phenomena, such as EIT, gain without inversion, and enhanced dispersion. Absorption/gain spectra possess extremely narrow sub-natural resonances. At strong enough probe field both the gain and the dispersion exhibit a flat-top behavior controlled by the relative phase of the probe and the coupling laser fields.


\end{document}